\pdfoutput=1
\documentclass[aps,pre,twocolumn,superscriptaddress]{revtex4-1}
\usepackage[english]{babel}
\usepackage{amsmath, amssymb, mathtools, bbm}
\usepackage{graphicx, tikz, epstopdf}

\usepackage[babel]{csquotes}
\usepackage{natbib}

\usepackage[colorlinks,linkcolor=blue,urlcolor=blue]{hyperref}

\addto\captionsenglish{\renewcommand{\figurename}{Fig.~\!}}
\newcommand{\Sec}{Sec.~\!}
\newcommand{\Secs}{Secs.~\!}
\newcommand{\Eq}{Eq.~\!}

\newcommand{\di}{\mathrm{d}}
\DeclareMathOperator{\Div}{div}

\begin{document}

\title{Coarse-graining via the fluctuation-dissipation theorem and large-deviation theory}
\author{Alberto Montefusco}
\affiliation{ETH Z\"{u}rich, Department of Materials, Polymer Physics, CH-8093 Z\"urich, Switzerland}
\author{Mark A. Peletier}
\affiliation{Technische Universiteit Eindhoven, Centre for Analysis, Scientific Computing and Applications, and Institute for Complex Molecular Systems (ICMS), 5600 MB Eindhoven, The Netherlands}
\author{Hans Christian \"{O}ttinger}
\affiliation{ETH Z\"{u}rich, Department of Materials, Polymer Physics, CH-8093 Z\"urich, Switzerland}
\date{\today}

\begin{abstract}
 The fluctuation-dissipation theorem is a central result in statistical mechanics and is usually formulated for systems described by diffusion processes. In this paper, we propose a generalization for a wider class of stochastic processes, namely the class of Markov processes that satisfy detailed balance and a large-deviation principle. The generalized fluctuation-dissipation theorem characterizes the deterministic limit of such a Markov process as a generalized gradient flow, a mathematical tool to model a purely irreversible dynamics via a dissipation potential and an entropy function: these are expressed in terms of the large-deviation dynamic rate function of the Markov process and its stationary distribution. We exploit the generalized fluctuation-dissipation theorem to develop a new method of coarse-graining and test it in the context of the passage from the diffusion in a double-well potential to the jump process that describes the simple reaction $A \rightleftarrows B$ (Kramers' escape problem).
\end{abstract}

\maketitle

\section{Coarse-graining via the fluctuation-dissipation theorem}
 A fluctuation-dissipation theorem of the second kind (FDT), according to the terminology of Kubo \cite{KTH85}, gives a one-to-one relationship between the noise and the friction properties of a diffusion process, and is formulated even far from equilibrium \cite{hcO05}. Our first aim is to generalize the FDT to the class of Markov processes. The motivation comes from a theory of coarse-graining, and our second aim is to generalize this, too.
 
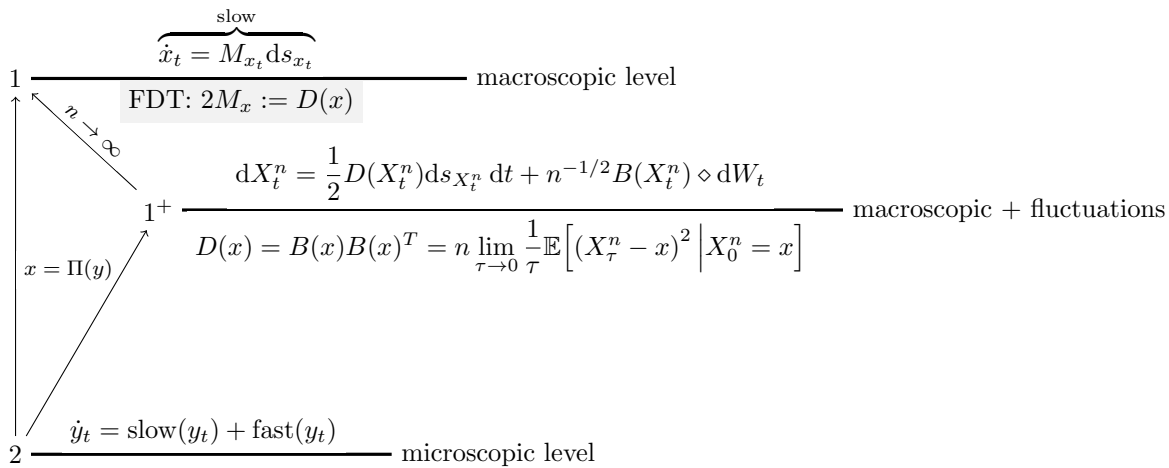
\begin{figure*}[t]
  \begin{tikzpicture}[node distance=1.75cm]
   \tikzstyle{every node}=[font=\normalsize, fill=white]
   \draw[very thick]									(0, 0) -- (5, 0)
    node[pos=.5, above]		{$\dot{y}_t = \operatorname{slow}(y_t) + \operatorname{fast}(y_t)$}
    node[at start]		(A)	{2}
    node[right] 			{microscopic level};
   \draw[very thick]									(0, 5) -- (6, 5)
    node[pos=.49, above]				{$\overbrace{\dot{x}_t = M_{x_t} \di s_{x_t}}^\text{slow}$}
    node[at start] 		(B) 			{1}
    node[right] 						{macroscopic level}
    node[midway, below, fill=gray!10]	{FDT: $2 M_x := D(x)$};
   \draw[very thick]									(1.9, 3.25) -- (11, 3.25)
    node[midway, above]					{$\di X^n_t = \dfrac{1}{2} D(X^n_t) \di s_{X^n_t} \, \di t + n^{-1/2} B(X^n_t) \diamond \di W_t$}
    node[at start] 		(C) 			{$1^+$}
    node[right] 						{macroscopic + fluctuations}
    node[midway, below]					{$D(x) = B(x) B(x)^T = n \lim\limits_{\tau \to 0} \dfrac{1}{\tau} \mathbb{E}\!\left[ \left( X^n_\tau - x \right)^2 \Big| X^n_0 = x \right]$};
   \draw[->]											(A) -- (B)
    node[pos=.49, right]		{\footnotesize{$x = \Pi(y)$}};
   \draw[->]											(A) -- (C);
   \draw[->]											(C) -- (B)
    node[midway, above, sloped]	{\footnotesize{$n \to \infty$}};
  \end{tikzpicture}
  \caption{Coarse-graining via the FDT. A microscopic level of description (2) is described by the variables $y$ and its dynamics may be decomposed into a slow and a fast component. Level $1^+$ is a stochastic approximation of level $2$: its dynamics is governed by an SDE that has $e^{n s}$ as a stationary distribution. At the macroscopic level (1), the reduced set of variables $x$ accounts for the slow dynamics only, and the dynamics is a gradient flow where the friction matrix is defined via the FDT. Hence, the friction matrix $M$ may be calculated via the evaluation of the diffusion coefficient $D$, which is done through the estimation of a second moment of the process: this procedure is the basis of the Green-Kubo relations.}
  \label{fig:FDT}
 \end{figure*}

 The goal of a theory of \emph{coarse-graining} is to derive more macroscopic from more microscopic models of a physical system. A general class of nonequilibrium-thermodynamical models can be written in the form of metriplectic systems \cite{pjM86} or the GENERIC \cite{GO97,OG97}, where the dynamics has the mathematical structure of a ``force'' times a ``phenomenological matrix'', which is directly related to the ``cometric'' field in the language of metriplectic systems, or the ``friction matrix'' in the language of the GENERIC. In this framework, the goal of coarse-graining is to resolve the mathematical structure of a macroscopic model in terms of the properties of the microscopic one, that is: (i) to find the thermodynamic potential~$s$ giving rise to the ``force'' and (ii) to compute the friction matrix~$M$. The classical setup is shown in \figurename\ref{fig:FDT}:
 \begin{itemize}
  \item A ``microscopic'' model (level~2), identified by the variables~$y$, shows a separation of time scales, such that the dynamics may be decomposed into a ``slow'' and a ``fast'' component. The first challenge is to identify a set of ``slow'' variables $x = \Pi(y)$ at the ``macroscopic'' level~1.
  \item One considers a stochastic extension of level 1, called ``level $1^+$'', which is assumed to be a diffusion process controlled by an extensive parameter~$n$. The noise term reproduces, in an approximate fashion, the fast dynamics that has been neglected in the transition  operated by the map~$\Pi$. In the deterministic limit, when the parameter~$n$ is very large, fluctuations vanish and we recover the macroscopic model at level~$1$.
  \item The entropy function is found by looking at the stationary distribution of the diffusion process, and the friction matrix is defined by the diffusion tensor through $2 M = D$. The latter definition expresses the FDT, which is a one-to-one relation between the diffusion matrix and the drift term of a diffusion equation, as a consequence of detailed balance with respect to an invariant measure of the Boltzmann form $e^{n s}$. This implies that the friction matrix can be computed by simulation of the microscopic dynamics and estimation of the second moments of the approximating stochastic process, which is the basis of \emph{Green-Kubo relations} \cite{KTH85,hcO05}.
 \end{itemize}
 This scheme has been advocated by various authors in the framework of the \emph{projection-operator} technique \cite{rwZ01,hG82,hcO05,pE04,EV02}, which allows us to derive the diffusion process at the level $1^+$ and the expression of the friction matrix in terms of the microscopic data in a formal manner. In the literature \cite{PS08,ZHS16} there exist other mathematical techniques that also produce diffusion processes as effective dynamics and lead to similar conclusions.
 
 Many fluctuating systems, however, are not well approximated by diffusion processes, but require more general Markov processes \cite{FVEW15,DLLN16}. A typical example is represented by chemical reactions, which are characterized by rare and large events and are described by Markov jump processes. These are substantially different from diffusion processes in that the latter evolve continuously in time as infinitesimally small movements in state space, while -- for the former -- it is always possible to find a time scale at which the dynamics appears as constituted of sudden jumps at discrete instants of time. As announced before, the second aim of this work is to extend the above scheme of coarse-graining to the setting where fluctuations are assumed of the form of Markov processes, and the generalized FDT serves exactly this purpose. In the context of Markov processes, the generalized FDT gives friction no longer in terms of a matrix, but in terms of a dissipation potential \cite{dgbE72,mG93,CV90}.
  
 The mathematical ingredients that we need are (i) \emph{generalized gradient flows} \cite{MPR14} or the (purely dissipative) GENERIC \cite{GO97} to formulate friction through dissipation potentials, and (ii) large-deviation theory \cite{hT09} to characterize fluctuations. Following \cite{MPR14}, we identify a correspondence between Markov processes describing fluctuations and the generalized gradient structures of their deterministic limit: we call this connection a generalized fluctuation-dissipation theorem of the second kind (generalized FDT). Thanks to the powerful tools of large deviations, in particular a numerical implementation of the Feng-Kurtz scheme \cite{FK06}, we propose a novel coarse-graining method that aims at computing dissipation potentials. We test our newly-devised method in the example of a very simple chemical reaction.
 
 All constructions are restricted to purely dissipative systems and Markov processes with detailed balance. The proper extension including reversible dynamics has not been established yet and represents one of the biggest open issues.
 
 Before giving our reformulation of an FDT, we introduce the concept of a generalized gradient flow in \Sec\ref{sec:GGF} and, in \Sec\ref{sec:LD}, we give a short intuitive account of large-deviation theory and its use in statistical mechanics. In \Sec\ref{sec:FDT}, we first illustrate the usual formulation of an FDT in the language of gradient flows and large deviations, thus providing the starting point for the intended generalization. Then, we formulate the generalized FDT for Markov processes and generalized gradient flows. In \Sec\ref{sec:CME}, we study an example in the context of chemical reactions with both analytic and numerical instruments. Our conclusions and perspectives may be found in \Sec\ref{sec:conclusions}.
 
\section{Dissipation: generalized gradient flows\label{sec:GGF}}
 A great deal of purely dissipative systems may be expressed in the mathematical language of \emph{generalized gradient flows}, a nonlinear generalization of \emph{gradient flows}. Such structures are recently experiencing growing interest among mathematicians and physicists. The mathematician uses them to prove existence and stability of solutions~\cite{AGS08}, or convergence of evolution equations in the limit of some parameter~\cite{aM16a}. The physicist benefits from geometric structures that express thermodynamics \cite{pjM86,GO97,OG97,gpB14}.
 
 A \emph{(standard) gradient structure} on the space $\mathcal{X}$ is a pair $\left( M, s \right)$, where $s \colon \mathcal{X} \to \mathbb{R}$ is a smooth function and $M$ is a symmetric and non-negative definite two-tensor field on $\mathcal{X}$, which is called a \emph{cometric} in \cite{pjM86} and a \emph{friction matrix} in \cite{hcO05}. A gradient structure induces the evolution equation
 \begin{equation}\label{GF}
  \dot{x}_t = M_{x_t} \di s_{x_t} \, ,
 \end{equation}
 which we call a \emph{(standard) gradient flow}.
 
 In the framework of nonequilibrium thermodynamics, gradient structures find inspiration in the linear relationships between fluxes and forces proposed by Onsager~\cite{lO31a}. When the friction matrix is directly related to the phenomenological (or Onsager) matrix, symmetry is a manifestation of Onsager's reciprocal relations, and positive semidefiniteness expresses the non-negativity of the entropy production. However, \Eq\eqref{GF} has been shown to accommodate several nonlinear force-flux relations, too, such as chemical reactions~\cite{aM11,hcO15}.
 
 \emph{Generalized gradient structures}~\cite{MPR14} were advanced in various contexts, prompted by considerations of mathematical structure~\cite{DGMT80,CV90,eDG93}, and geometrical and physical meaning~\cite{mG93,MPR15}, and chemical reactions are the prototypical example \cite{mG12}. As outlined in \Sec\ref{sec:CME}, such structures find a further motivation in the light of the generalized FDT, where they arise from the properties of some underlying level of descriptions through the form of the fluctuations.
 
 The generalization is based on the observation that \Eq\eqref{GF} can also be written as
 \begin{equation}\label{GGF}
  \boxed{\dot{x}_t = \partial_\xi \Psi^*_{x_t}\!(\di s_x)} \, ,
 \end{equation}
 where the \emph{dissipation potentials} \footnote{The quadratic form $v \cdot M_x^{-1} v$, $M_x$ possibly being degenerate, does not exist in the ordinary sense. In the context of convex analysis, however, it should be interpreted as $\xi \cdot M_x \xi$ if $v \in \mathcal{R}(M_x)$ and $M_x \xi = v$, and $+ \infty$ otherwise, where $\mathcal{R}$ is the range.}
 \begin{equation*}
  \Psi_x(v) := \dfrac{1}{2} v \cdot M_x^{-1} v \quad \text{ and } \quad \Psi^*_x(v) := \dfrac{1}{2} \xi \cdot M_x \xi
 \end{equation*}
 are dual to each other in the sense of Legendre-Fenchel transforms \cite{rtR70,hT14}:
 \begin{subequations}
  \begin{align}\label{dual}
   \Psi^*_x(\xi) &= \sup\limits_v \left[ \xi \cdot v - \Psi_x(v) \right] \qquad \text{and} \\
   \Psi_x(v) 	&= \sup\limits_\xi \left[ \xi \cdot v - \Psi^*_x(\xi) \right] \, .
  \end{align}
 \end{subequations}
 
 A \emph{generalized gradient flow} is of the form \eqref{GGF}, but with $\Psi$ and $\Psi^*$ not necessarily quadratic. More precisely,
 \begin{center}
  \noindent\fbox{\parbox{.92\linewidth}{the pair $\left( \Psi, s \right)$ is a \emph{generalized gradient structure} (GGS) on $\mathcal{X}$ if, for all $x \in \mathcal{X}$,
   \begin{enumerate}
  	\item $\Psi_x(v)$ is convex in the variable $v$,
  	\item $\Psi_x(0) = 0$,
  	\item $\min\limits_v \Psi_x(v) = 0$.
   \end{enumerate}\vspace{-3.5mm}}}
 \end{center}
 A dissipation potential is called \emph{symmetric} if $\Psi_x(v)=\Psi_x(-v)$ for all $(x, v)$. It can be verified that $\Psi^*$ inherits exactly the same properties and that, when symmetry is satisfied, properties 1 and 2 imply 3.
 
 There is another formulation of the generalized gradient flow \eqref{GGF}, as a minimization problem, that is particularly useful to our work. Indeed, from \Eq\eqref{dual}, the \emph{Young-Fenchel inequality} follows,
 \begin{equation}
  \Psi_{x}(v) + \Psi^*_{x}(\xi) - \xi \cdot v \geq 0 \, ,
 \end{equation}
 and this inequality holds with an equal sign when $v$ is a solution $\bar{v}(\xi)$ of the maximization problem in \Eq\eqref{dual}. Then, let us define the function
 \begin{equation}\label{FFunction}
  \mathcal{F}(x, v) := \Psi_{x}(v) + \Psi^*_{x}\big(\di s_x\big) - \di s_x \cdot v \, ,
 \end{equation}
 which is convex in its second argument and, by the Fenchel-Young inequality, is always non-negative.
 \begin{center}
  \noindent\fbox{\parbox{.92\linewidth}{The generalized gradient flow \eqref{GGF} has the equivalent characterization
   \begin{equation}\label{evolution}
   	\mathcal{F}(x_t, \dot{x}_t) = 0 \, ,
   \end{equation}\vspace{-7mm}}}
 \end{center} 
 namely, the trajectories $x \colon [0, T] \to \mathcal{X}$ minimize $\mathcal{F}$ at value zero. This formulation also extends \Eq\eqref{GGF} to the case of non-differentiable dissipation potentials.
 
 Now we are able to make two important remarks on the definition of a GGS: first, conditions 2 and 3 imply that stationary points of $s$ are also stationary solutions of the evolution equation \eqref{evolution}; furthermore, $s$ is a Lyapunov function of the evolution because, along a solution $x \colon [0, T] \to \mathcal{X}$ of \Eq\eqref{evolution},
 \begin{equation*}
 \dfrac{\di (s \circ x)(t)}{\di t} = \di s_{x_t} \cdot \dot{x}_t = \Psi_{x_t}(\dot{x}_t) + \Psi^*_{x_t}\!\big(\di s_{x_t}\big) \geq 0 \, ,
 \end{equation*}
 since $\Psi$ and $\Psi^*$ are both non-negative. These two features constitute essential reasons for the use of such structure in thermodynamics, where the driving function $s$ is identified with the thermodynamic entropy and the dissipation potential provides the relationship between nonequilibrium forces and fluxes.
 
\section{Fluctuations: the theory of large deviations\label{sec:LD}}
 In the previous section we have introduced a mathematical structure that expresses dissipation. The trajectories of the system minimize the function $\mathcal{F}$ at every instant of time. In this section, we study fluctuations around trajectories, and realize that a similar minimization feature arises in the framework of large-deviation theory.
  
 \subsection{Three notions of convergence}
 In order to describe fluctuations, we use random variables and stochastic processes on $\mathcal{X}$. In particular, since we are interested in systems typical of statistical mechanics, characterized by many degrees of freedom, we consider sequences of random variables and stochastic processes indexed by some large parameter $n$. We will use the symbol $X^n$ to denote both sequences of random variables and of stochastic processes: the latter, indeed, can be seen as random variables taking values in some space of curves in $\mathcal{X}$.
 
 In the limit $n \to \infty$, different limit theorems, corresponding to distinct notions of convergence, come into play.
 
 \paragraph{The law of large numbers.}
  If the probability distribution concentrates onto a single point $z \in \mathcal{X}$, we say that we have a \emph{law of large numbers}, that is
  \begin{equation}
   X^n \to z \qquad \text{almost surely as } n \to \infty \, .
  \end{equation}
  The point $z$ is called the \emph{deterministic limit} of $X^n$. When $X^n$ is a stochastic process, $z$ is also called the \emph{deterministic evolution}.
 
 \paragraph{The central limit theorem.}
  Now, let us suppose we have a law of large numbers, consider a \emph{fluctuation}
  \begin{equation*}
   X^n - z \, ,
  \end{equation*}
  and rescale it by a factor $n^{1/2}$:
  \begin{equation}\label{CLT}
   W^n := n^{1/2} \left( X^n - z \right) \, .
  \end{equation}
  
  When a central limit theorem holds, it tells us that $W^n$ converges in distribution to a normal random variable. This represents a first characterization of fluctuations around the deterministic limit, one that describes small, relatively probable deviations.
  
 \paragraph{Large deviations.}
  A further characterization is provided by the theory of \emph{large deviations}, which studies untypical occurrences contained in the tail of the distribution. As $n \to \infty$ the probability of such an occurrence tends to zero, and large-deviation theory searches for a decay of the form
  \begin{equation}\label{LDP}
   \mathbb{P} \!\left( X^n \approx x \right) \asymp e^{- n I(x)} \, ,
  \end{equation}
  which is called a \emph{large deviation principle} (LDP) and reads: the probability of the random variable $X^n$ to be close (here vaguely denoted by ``$\approx$'' ) to some $x \in \mathcal{X}$ decays exponentially with a rate given by $n$ times a \emph{rate function}. A rate function is a lower-semicontinuous function $I \colon \mathcal{X} \to [0, \infty]$.
  
  The symbol ``$\asymp$'' defines the notion of large-deviation convergence for random variables: a detailed account of the mathematical theory can be found in the book \cite{DZ10}, and an excellent presentation for physicists is contained in the review article \cite{hT09}. Given a certain physical setup represented by a sequence of random variables, the convergence in the large-deviation sense is a result that needs to be proven with the appropriate mathematical tools. We give a short account of one of these tools at the end of this section and use it in our numerical experiment.
  
  We remark that, whenever the rate function has a unique minimum $0$ at $z \in \mathcal{X}$, the LDP \eqref{LDP} automatically yields a (strong) law of large numbers. Indeed, this property implies that
  \begin{equation*}
   X^n \to z \qquad \text{almost surely as } n \to \infty \, .
  \end{equation*}
  Moreover, in good cases, the quadratic approximation of the rate function around $z$ reproduces the central limit theorem \cite{wB93}.
  
  Similar definitions exist for stochastic processes. Now, consider curves and stochastic processes in $\mathcal{X}$. Then, a large-deviation principle reads
  \begin{equation*}
   \mathbb{P} \!\left( \left. X^n_t \right\rvert_{[0, T]} \approx \left. x_t \right\rvert_{[0, T]} \right) \asymp e^{- n I_{[0, T]}(x)} \qquad \text{as } n \to \infty \, ,
  \end{equation*}
  which means: the probability of the stochastic process to be close to some curve $x \colon [0, T] \to \mathcal{X}$ decays exponentially with a rate given by $n$ times a rate function, which is again a non-negative lower-semicontinuous function on the set of curves in $\mathcal{X}$. If the rate function has a unique minimum 0, we may again identify an element in the space of curves that is the deterministic limit of the stochastic process.
  
  \subsection{Large deviations and statistical mechanics}
  The language of large-deviation theory is central to statistical mechanics. As recognized in \cite{hT09}, any statement that connects probabilities of microstates to a specific thermodynamic potential is an LDP. For instance, the definition of Boltzmann's entropy
  \begin{equation}\label{Boltzmann}
   S^n_\text{eq}(u) := k_B \ln \mathbb{P}(H^n/n \approx u) \, ,
  \end{equation}
  where $H^n$ is the total energy and $u$ is a possible realization of the energy per particle $H^n/n$, leads to an LDP in the following sense. Let us ask whether the specific entropy
  \begin{equation*}
   s_\text{eq}(u) := \lim\limits_{n \to \infty} \dfrac{S^n_\text{eq}(u)}{n}
  \end{equation*}
  exists. If we have \Eq\eqref{Boltzmann}, then
  \begin{equation*}
   s_\text{eq}(u) = \lim\limits_{n \to \infty} \dfrac{1}{n} k_B \ln \mathbb{P}(H^n/n \approx u) \, ,
  \end{equation*}
  or
  \begin{equation*}
   \mathbb{P}(H^n/n \approx u) \asymp e^{n s_\text{eq}(u)} \, .
  \end{equation*}
  The LDP tells us that the entropy $S^n_\text{eq}$ becomes extensive ($S^n_\text{eq} \simeq n s_\text{eq}$) in the limit of many degrees of freedom.
  
  Similar relations hold for other thermodynamic potentials, and the structure of Legendre-Fenchel transforms among them completely reflects the same structure arising in large-deviation theory \cite{hT09}.
  
  The equilibrium states are defined as the most probable states in the limit $n \to \infty$, namely, the minimizers of the rate function. The rate function describes the fluctuations around the equilibrium states.
  
  What happens in nonequilibrium statistical mechanics? Can similar statements be established? In the present paper, we will give a characterization of macroscopic dynamics as minimizers of large-deviation rate functions for stochastic processes. The rate functions, then, will describe the fluctuation paths around the deterministic evolution. Moreover, the characterization will provide the deterministic dynamics with a precise structure.
  
 \subsection{The Feng-Kurtz method}\label{Feng-Kurtz}
  We conclude this section by a concise description of the Feng-Kurtz method \cite{FK06}, which represents both a useful tool to derive explicit expressions for the rate functions and a well-developed theory to rigorously prove corresponding mathematical statements.
  
  Given a sequence of time-homogeneous Markov processes with infinitesimal generator
  \begin{equation}
   (\mathcal{Q}^n f)(x) := \lim\limits_{t \to 0} \dfrac{\mathbb{E}\!\left[ f(X^n_t) \big| X^n_0 = x \right] - f(x)}{t} \, ,
  \end{equation}
  we want to establish an LDP. To start with, we define the \emph{nonlinear} or \emph{Fleming generator} \cite{whF78}
  \begin{equation}
   (H_n f)(x) := \dfrac{1}{n} e^{-n f(x)} (\mathcal{Q}^n e^{n f})(x)
  \end{equation}
  and search for a limit
  \begin{equation*}
   H_n \rightarrow H
  \end{equation*}
  in some operator sense that we don't define here. The convergence of the nonlinear generator implies the existence of a large-deviation principle \cite{FK06}, and the limit~$H$ leads to a characterization of the rate function, as follows.
  
  For a Markov process, the expression
  \begin{equation*}
   (H f)(x)
  \end{equation*}
  depends on the function $f$ only through its first derivative. We can therefore define the \emph{Hamiltonian}
  \begin{equation}
   \mathcal{H}(x, \di f(x)) := (H f)(x) \, ,
  \end{equation}
  and compute its Legendre-Fenchel transform
  \begin{equation}
   \mathcal{L}(x, v) = \sup\limits_\xi \left[ \xi \cdot v - \mathcal{H}(x, \xi) \right] \, ,
  \end{equation}
  which we call the \emph{Lagrangian}. Then, the rate function has the form
  \begin{equation*}
   I_{[0, T]}(x) = I_0(x_0) + \int_0^T \mathcal{L}(x_t, \dot{x}_t) \, \di t \, ,
  \end{equation*}
  where $I_0$ is the rate function for the initial state $X^n_0$. Since $I_0$ plays no role in the following, we suppose $X^n_0$ is chosen deterministically and write concisely
  \begin{equation}\label{rateFunction}
   I_{[0, T]}(x) = \int_0^T \mathcal{L}(x_t, \dot{x}_t) \, \di t \, .
  \end{equation}
  
  As we may expect for time-homogeneous Markov processes, information is contained in a function of just two variables (the Lagrangian), which expresses the fact that -- for such processes -- information is local in time. We will take advantage of this property in our numerical simulations of \Sec\ref{numerics}. For a concrete example of the analytical procedure, see \Sec\ref{largeDeviationsCME}.
  
  The form of the rate function \eqref{rateFunction} presents the following additional feature: when a law of numbers holds, the deterministic evolution must minimize the Lagrangian at every instant of time. This establishes a parallel with the discussion of generalized gradient flows that we will elaborate in our generalization of the FDT.

\section{The generalized fluctuation-dissipation theorem\label{sec:FDT}}
 In \Secs\ref{sec:GGF}-\ref{sec:LD} we have introduced the mathematical ingredients that we need to express dissipation and fluctuations and we have recognized a similarity in that both ingredients are characterized by a minimization of a function: $\mathcal{F}$ in the case of generalized gradient flows, $\mathcal{L}$ in the case of LDPs.
 
 Now we are ready to address the first aim of the paper: the generalization of the FDT. Given a sequence of stochastic processes that approximate some microscopic dynamics, what is the proper dissipative structure of its deterministic limit, namely, the macroscopic dynamics?
 
 As a first step, we introduce the classical formulation of the FDT for diffusion processes. Then, we translate it into the language of large deviations and GGSs. Finally, we use this translation to state the generalization for Markov processes.
 
 \subsection{FDT for diffusion processes and gradient flows}\label{FDT-GF}
  According to the coarse-graining procedure depicted in \figurename\ref{fig:FDT}, for a wide class of systems the correct description of fluctuations is given in terms of a diffusion process. This happens when the dynamics of the macroscopic variables~$X^n$ is the sum of the short-time correlated interactions of many microscopic particles, such that the fluctuations are modeled as a Gaussian white noise \cite{OM53}. The governing equation is the stochastic differential equation (SDE)
  \begin{equation}\label{diffusion}
   \di X^n_t = A(X^n_t) \, \di t + n^{-1/2} B(X^n_t) \diamond \di W_t \, ,
  \end{equation}
  where the expressions for the functions $A$ (called \emph{drift}) and $B$ (the \emph{noise intensity matrix}) need to be specified. This equation can be formally derived by projection-operator techniques, which also give the expressions for $A$ and $B$ in terms of the microscopic data \cite{pE04,EV02}. The symbol $\diamond$ stands for the \emph{kinetic} or \emph{Klimontovich} interpretation for the noise \cite{HO98}, which -- in It\^{o} form -- results in the SDE
  \begin{gather*}
   \di X^n_t = \left[ A(X^n_t) + \dfrac{1}{2 n} \Div D(X^n_t) \right] \di t + \dfrac{B(X^n_t)}{\sqrt{n}} \, \di W_t \, , \\
   \text{with } D(x) := B(x) B(x)^T \, .
  \end{gather*}
  
  Now, suppose that this process has a stationary distribution of the Boltzmann type \footnote{The choice of the kinetic interpretation for the noise has been made to let this stationary distribution be such for any value of $n$. To support the following arguments, however, we need a weaker condition: that the distribution satisfies the LDP $\pi^n \asymp e^{n s}$. This fact makes the choice of interpretation for the noise irrelevant, since the correction in the drift vanishes in the limit $n \to \infty$.},
  \begin{equation}\label{eq}
   \pi^n_x = e^{n s_x} \, ,
  \end{equation}
  and, in addition, is in detailed balance with respect to it. This means that the generator of the process \eqref{diffusion},
  \begin{equation*}
  (Q^n f)(x) = A(x) \cdot \di f(x) + \dfrac{1}{2 n} \Div \!\big[ D_x \di f(x) \big] \, ,
  \end{equation*}
  is self-adjoint with respect to the measure \eqref{eq}, viz.,
  \begin{equation*}
  \int_\mathcal{X} f(x) \, (Q^n g)(x) \, e^{n s_x} \, \di x = \int_\mathcal{X} g(x) \, (Q^n f)(x) \, e^{n s_x} \, \di x
  \end{equation*}
  for all functions $f$ and $g$ in a proper class. Then, the SDE \eqref{diffusion} must have the form \cite[\Sec 6.3.5]{cG09}
  \begin{equation}\label{diffDB}
   \di X^n_t = \dfrac{1}{2} D(X^n_t) \di s_{X^n_t} \, \di t + n^{-1/2} B(X^n_t) \diamond \di W_t \, .
  \end{equation}
  We call this result the \emph{classical FDT}: \emph{given detailed balance, the drift term has the structure of a gradient flow, where the driving function is related to the logarithm of the stationary distribution of the process and the linear operator is (a half of) the diffusion tensor}. Of course, the deterministic limit of \eqref{diffDB}, which describes the macroscopic evolution, is given by the drift:
  \begin{equation*}
   \dot{x}_t = \dfrac{1}{2} D(x) \di s_x \, .
  \end{equation*}
  Therefore, we \emph{define} the friction matrix to be equal to a half of the diffusion tensor,
  \begin{equation}\label{FDT}
   \boxed{M_x := \dfrac{1}{2} D(x)} \, ,
  \end{equation}
  and the macroscopic evolution equation is the gradient flow
  \begin{equation}\label{deterministicGF}
   \dot{x}_t = M_{x_t} \di s_{x_t} \, .
  \end{equation}
    
 \subsection{Reformulation of the FDT via large deviations}
  As already noticed by Onsager and Machlup \cite{OM53}, the path measure of \Eq\eqref{diffDB} has a special form that encodes all data about the structure of the deterministic limit. In the language and notation of the present paper, the path measure satisfies an LDP
  \begin{equation*}
   \mathbb{P} \!\left( \left. X^n_t \right\rvert_{[0, T]} \approx \left. x_t \right\rvert_{[0, T]} \right) \asymp e^{- n I_{[0, T]}(x)}
  \end{equation*}
  with
  \begin{equation*}
   I_{[0, T]}(x) = \int_0^T \mathcal{L}(x_t, \dot{x}_t) \, \di t
  \end{equation*}
  and
  \begin{equation}\label{LagrangianDiff}
   \mathcal{L}(x, v) = \dfrac{1}{2} \left( v - \dfrac{D(x)}{2} \di s_x \right) \cdot D(x)^{-1} \left( v - \dfrac{D(x)}{2} \di s_x \right) \, .
  \end{equation}
  A corresponding result can be proven rigorously for any SDE of the form \eqref{diffusion} \cite{FW98} and can also be formally obtained by using the Feng-Kurtz method explained in \Sec\ref{Feng-Kurtz}.
  
  The Lagrangian \eqref{LagrangianDiff} is (a half of) the function that generates the standard gradient structure $\left( D/2, s \right)$: this is exactly equivalent, and represents nothing but a reformulation, of the classical FDT. Therefore, we define the function $\mathcal{F}$, which drives the macroscopic gradient flow, to be
  \begin{equation}
   \mathcal{F} := 2 \mathcal{L} \, .
  \end{equation}
  Our generalization will elaborate exactly on this observation.
  
 \subsection{Generalized FDT for Markov processes with detailed balance}
  We have just seen that the large-deviation Lagrangian of a diffusion process with detailed balance is the $\mathcal{F}$-function of a standard gradient flow, the entropy being related to the logarithm of the stationary distribution. This is a first realization of the parallel between generalized gradient flows and large-deviation Lagrangians that we suggested before. The parallel is even more general when we extend the class of processes to Markov processes with detailed balance. 
  
  Assume that a sequence of Markov processes $X^n$ satisfies the following conditions:
  \begin{enumerate}
   \item convergence to a deterministic curve that solves
    \begin{equation*}
     \dot{z}_t = \mathcal{A}(z_t) \, ;
    \end{equation*}
   \item the LDP
    \begin{equation*}
     \mathbb{P} \!\left( \left. X^n_t \right\rvert_{[0, T]} \approx \left. x_t \right\rvert_{[0, T]} \right) \asymp e^{- n I_{[0, T]}(x)}
    \end{equation*}
    with rate function
    \begin{equation}\label{FDT-rateFunction}
     I_{[0, T]}(x) = \int_0^T \mathcal{L}(x_t, \dot{x}_t) \, \di t
    \end{equation}
    and $\mathcal{L}(x, v)$ convex in $v$;
   \item detailed balance with respect to the stationary distribution
    \begin{equation}
     \pi^n_x \asymp e^{n s_x} \, .
    \end{equation}
  \end{enumerate}
  Then, it is proven in \cite{MPR14} that \emph{the large-deviation Lagrangian generates a generalized gradient structure (with symmetric dissipation potential) for the deterministic limit}. We call this statement a \emph{generalized FDT}. As a consequence, we define the $\mathcal{F}$-function of the deterministic limit by
  \begin{equation}\label{genFDT}
   \boxed{\mathcal{F} := 2 \mathcal{L}} \, . \\
  \end{equation}
  The function $\mathcal{F}$ has of course the form
  \begin{equation*}
   \mathcal{F}(x, v) = \Psi_x(v) + \Psi^*_x(\di s_x) - \di s_x \cdot v
  \end{equation*}
  and \emph{the dissipation potentials are symmetric}.
  
  Strictly speaking, detailed balance is sufficient but not necessary to obtain a GGS as defined in \Sec\ref{sec:GGF} \cite{MPR14}: it would suffice that
  \begin{equation*}
   2 \partial_v \mathcal{L}(x, 0) = - \di s_x \, ,
  \end{equation*}
  whereas detailed balance implies that
  \begin{equation*}
   \mathcal{L}(x, v) - \mathcal{L}(x, -v) = - \di s_x \cdot v \, ,
  \end{equation*}
  which is a stronger statement. From the latter condition, one deduces the symmetry of the dissipation potentials, which -- in turn -- implies the properties 2 and 3 of the definition of a dissipation potential.
  
 \subsection{Generalized coarse-graining procedure}
\begin{figure*}[t]
   \begin{tikzpicture}[node distance=1.75cm]
    \tikzstyle{every node}=[font=\normalsize, fill=white]
    \draw[very thick]									(0, 0) -- (5, 0)
 	 node[pos=.5, above]	{$\dot{y}_t = \operatorname{slow\&fast}(y_t)$}
 	 node[at start]		(A)	{2}
 	 node[right] 			{microscopic level};
    \draw[very thick]									(0, 5) -- (6, 5)
 	 node[pos=.5, above]				{$\overbrace{\mathcal{F}(x_t, \dot{x}_t) = 0}^\text{slow}$}
 	 node[at start] 		(B) 		{1}
 	 node[right] 						{macroscopic level}
 	 node[midway, below, fill=gray!10]	{FDT: $\mathcal{F} := 2 \mathcal{L}$};
    \draw[very thick]									(1.9, 3.25) -- (13, 3.25)
 	 node[midway, above, align=center]	{sequence of Markov processes with $\mathbb{P}\!\left( X^n \approx x \right) \asymp e^{-n \int_{0}^{T} \mathcal{L}(x_t, \dot{x}_t) \, \di t}$ \\ and detailed balance with respect to $\pi^n \asymp e^{n s}$}
 	 node[at start] 		(C) 			{$1^+$}
 	 node[right] 						{macroscopic + fluctuations};
    \draw[->]											(A) -- (B)
 	 node[pos=.49, right]		{\footnotesize{$x = \Pi(y)$}};
    \draw[->]											(A) -- (C);
    \draw[->]											(C) -- (B)
 	 node[midway, above, sloped]	{\scriptsize{$n \to \infty$}};
   \end{tikzpicture}
   \caption{Coarse-graining for Markov processes via the generalized FDT. A microscopic level of description (2) is described by the variables~$y$ and its dynamics contains both fast and slow behaviors. Level~$1^+$ is a stochastic approximation of level~$2$: its dynamics is a sequence of Markov processes with an LDP. At the macroscopic level (1), the reduced set of variables~$x$ accounts for the ``slow'' dynamics only, and the dynamics is a generalized gradient flow where the function~$\mathcal{F}$ is defined via the generalized FDT. Hence, the driving function~$\mathcal{F}$ may be calculated via the evaluation of the Lagrangian~$\mathcal{L}$, which is done though the estimation of the nonlinear generator of the process.}
   \label{fig:GFDT}
  \end{figure*}
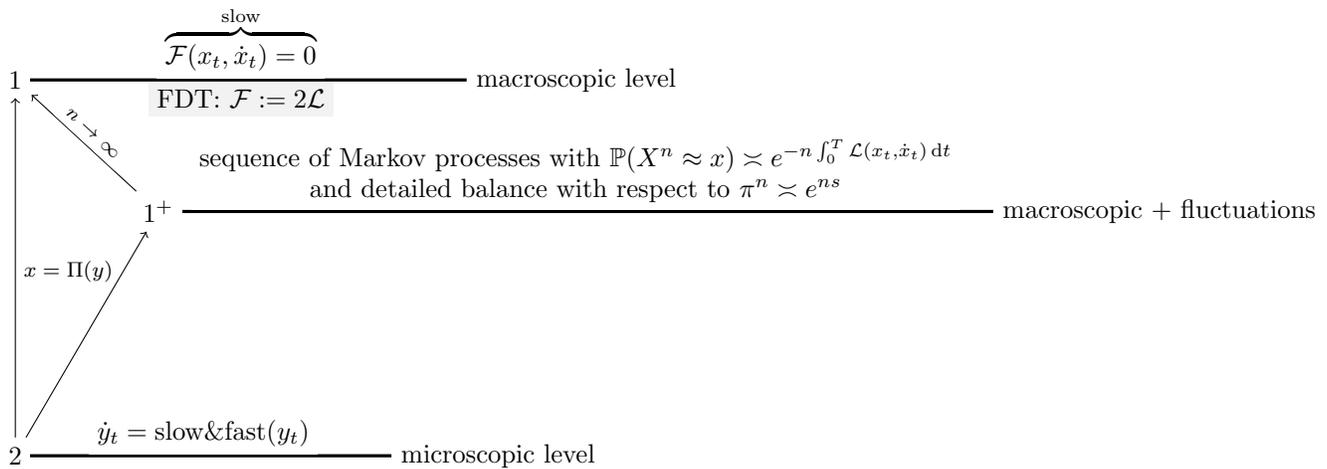

  A major consequence of the generalized FDT \eqref{genFDT} is an extension of the coarse-graining procedure described in the introduction. In \figurename\ref{fig:GFDT} we illustrate the effect of this generalization on the structure of \figurename\ref{fig:FDT}: whereas the method of \figurename\ref{fig:FDT} allowed us to handle microscopic dynamics that are well-approximated by diffusion processes, the scheme of \figurename\ref{fig:GFDT} extends this to microscopic systems whose fluctuations have the form of (sequences of) more general Markov processes. The typical examples that did not fit the scheme of \figurename\ref{fig:FDT} are systems characterized by rare events, which are mathematically described by Markov jump processes. The procedure outlined here constitutes a novel approach in the field of rare-event simulations.
  
  The generic setup is the following:
  \begin{itemize}
   \item We simulate the microscopic dynamics with variables~$y$ and we observe the dynamics of the coarse-grained variables $x = \Pi(y)$.
   \item In terms of the variables~$x$, the entropy~$s$ is the rate function of the stationary distribution.
   \item The function $\mathcal{F}$ that drives the macroscopic generalized gradient flow is two times the Lagrangian of the stochastic process, which we choose to calculate via a numerical implementation of the Feng-Kurtz method.
  \end{itemize}
  Once we have~$\mathcal{F}$, it is easy to get the dissipation potential upon the observation
  \begin{equation*}
   \mathcal{F}(x, 0) = \Psi_x(0) + \Psi^*_x(\di s) - \di s_x \cdot 0 = \Psi^*_x(\di s) \, ,
  \end{equation*}
  whence
  \begin{equation}
   \Psi_x(v) = \mathcal{F}(x, v) - \mathcal{F}(x, 0) + \di s_x \cdot v \, .
  \end{equation}
  We show an example of this procedure in the following section, where we compare the results with the exact ones predicted from theory.
  
\section{Example: chemical reactions\label{sec:CME}}
 In this section we describe an application of the generalized FDT in the context of chemical kinetics. Chemical reactions, together with all systems that exhibit rare-event features, are the archetypal example where Markov jump processes provide the right form of the fluctuations. The use of Green-Kubo schemes associated with diffusion processes (cf.~\figurename\ref{fig:FDT}) would lead to inaccurate or wrong result, as we show in \Sec\ref{Green-Kubo}.
 
 A typical microscopic model for chemical reactions is a diffusive dynamics in a potential landscape, which represents the effective interactions of all constituents of a mixture in the configuration space. If the minima of the landscape are separated by sufficiently high barriers with respect to the amplitude of the noise, the effective dynamics, at large time scales, is well approximated by a Markov jump process between the minima. The simplest realization of this scheme is Kramers' escape problem \cite{haK40}, which we analyze in detail.
 
 We introduce the model from the standpoints of coarse-graining and the framework of the present paper and, in the last subsection, we propose a numerical strategy that follows the scheme of \figurename\ref{fig:GFDT}.
 
 \subsection{A multi-scale view: the levels of description}\label{A}
 
  \subsubsection{Level 2: diffusion in a double-well potential}
   \begin{figure}[h]
   	\centering
   	\begin{tikzpicture}[scale=1.85]
   	 \draw[->] 	(-1.5,0) -- (1.5,0)	node[right] 			{$q$};
   	 \foreach \x/\xtext in {0/0, -1/-1, 1/1}
   	  \draw[shift={(\x,0)}] (0pt,2pt) -- (0pt,-2pt) node[below] {$\xtext$};
   	 \draw[->] 	(0,0) -- (0,2) 		node[right, pos=.55] 	{$1$};
   	 \draw[domain=-1.5:1.5,smooth,variable=\x,black,thick] plot ({\x},{\x*\x*\x*\x - 2*\x*\x + 1}) node[above] {$V(q)$};
   	\end{tikzpicture}
   	\caption{The double-well potential $V$ that we use in the numerical experiment. The regions $A = (-\infty, 0)$ and $B = (0, \infty)$ are representative of the two chemical states.}
   	\label{fig:potential}
   \end{figure}
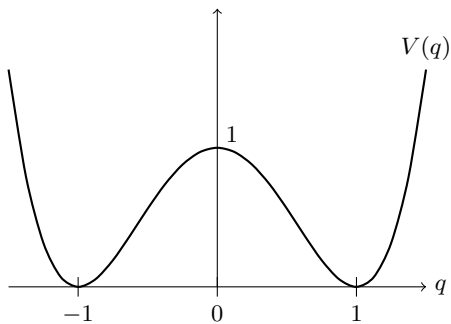
   We consider the overdamped Langevin dynamics of a Brownian particle in the potential of \figurename\ref{fig:potential}, whose two wells represent the two chemical states $A$ and $B$. The motion is governed by the SDE
   \begin{equation}
    \di Q^\epsilon_t = - \dfrac{1}{\gamma} V'(Q^\epsilon_t) \, \di t + \sqrt{\dfrac{k_B T}{\gamma}} \, \di W_t \, ,
   \end{equation}
   where $\gamma$ is a friction coefficient of unit $\textrm{kg}/\textrm{s}$. For simplicity, we put $\gamma = 1$: a different value would simply require a rescaling of time. Since we are interested in the low-temperature limit, we set $k_B T =: 2 \epsilon$ and write
   \begin{equation}\label{Kramers-SDE}
    \di Q^\epsilon_t = - V'(Q^\epsilon_t) \, \di t + \sqrt{2 \epsilon} \, \di W_t \, .
   \end{equation}
   Then, we take $n$ independent particles $Q^{\epsilon, i}$ in the same potential: these are the microscopic variables $y$.
  
  \subsubsection{Level $1^+$: the chemical master equation}
   Denoting by $\mathbbm{1}_J$ the indicator function of the set $J$, we consider the stochastic process
   \begin{equation}\label{empiricalProcess}
    X^{n, \epsilon} = \Pi(Q^{\epsilon, 1}, \ldots, Q^{\epsilon, n}) := \dfrac{1}{n} \sum_{i = 1}^{n} \mathbbm{1}_A(Q^{\epsilon, i}) \, ,
   \end{equation}
   which is the basic object we are interested in. It keeps track of the concentration of the particles that, at time $t$, are in the well $A$; namely, it is a rational number $x$ in the set
   \begin{equation*}
    \mathcal{X}^n = \left\{0, \dfrac{1}{n}, \dfrac{2}{n}, \dfrac{3}{n}, \ldots, \dfrac{n-1}{n}, 1 \right\} \subset [0, 1] \eqqcolon \mathcal{X} \, .
   \end{equation*}
   The concentration of $B$ is $1 - X^{\epsilon, n}$, of course. 
   
   As one can infer from \figurename~\ref{fig:potential}, when $\epsilon$ is sufficiently small, the process concentrates onto the minima of the potential, and moving between the two wells is a relatively rare event. In this regime, the system is characterized by two neatly separated time scales: the equilibration time $t_2$ of the particles in the wells, and the escape time $t_1$ from the wells \cite{DLLN16}, with $t_1 \gg t_2$. During the equilibration time $t_2$, a particle gets locally equilibrated and forgets where it was before the last jump event. In the limit $\epsilon \to 0$, the jumps become infinitely rare and, to obtain a non-trivial dynamics, one has to rescale time appropriately. By this means, the process \eqref{empiricalProcess} converges to a Markov jump process on the two chemical states,
   \begin{equation}\label{measureCME}
    X^{n, \epsilon} \to X^n \, .
   \end{equation}
   The evolution equation satisfied by the law of the process \eqref{measureCME} is the simplest realization of the \emph{chemical master equation} (CME) \cite{dtG92}. We think of the CME as approximating the $n$-particle diffusion process for small enough~$\epsilon$.
   
   The transition rates are given by the formulas
   \begin{equation*}
    \begin{cases}
     \mathcal{Q}\!\left( x \to x + \dfrac{1}{n} \right) = k \, n (1 - x) \\
     \mathcal{Q}\!\left( x \to x - \dfrac{1}{n} \right) = k \, n x
    \end{cases},
   \end{equation*}
   in terms of the \emph{reaction constant} $k := t_1^{-1}$, which means that the generator is the following linear operator on continuous bounded functions on $\mathcal{X}^n$:
   \begin{align}\label{generator}
    (\mathcal{Q}^n f)(x) &= n k (1 - x) \left[ f\!\left(x + \dfrac{1}{n}\right) - f(x) \right] + \nonumber \\
    &- n k x \left[ f(x) - f\!\left(x - \dfrac{1}{n}\right) \right] \, .
   \end{align}
  
  \subsubsection{Level 1: the reaction rate equation}
   From \Eq\eqref{generator}, we may immediately infer the deterministic dynamics. Indeed, as $n \to \infty$,
   \begin{equation}
    (\mathcal{Q}^n f)(x) \to (\mathcal{Q} f)(x) = \left[ k (1 - x) - k x \right] f'(x) \, ,
   \end{equation}
   which is a process with pure drift: it describes the deterministic \emph{reaction rate equation} (RRE) \footnote{also called \emph{law-of-mass-action dynamics} or \emph{Guldberg-Waage dynamics} \cite{GW867,sS87,mG12}}
   \begin{equation}\label{RRE}
    \dot{x}_t = k (1 - 2 x_t) \, ,
   \end{equation}
   with $x \in \mathcal{X} = [0, 1]$.
   
 \subsection{Large deviations for the chemical master equation}\label{largeDeviationsCME}
  The convergence of the CME to the RRE is characterized by the large-deviation principle
  \begin{equation*}
   \mathbb{P} \!\left( \!\left. X^n_t \right\rvert_{[0, T]} \approx \left. x_t \right\rvert_{[0, T]} \right) \asymp e^{- n I_{[0, T]}(x)} \qquad \text{as } n \to \infty \, ,
  \end{equation*}
  where
  \begin{equation*}
   I_{[0, T]}(x) = \int_0^T \mathcal{L}(x_t, \dot{x}_t) \, \di t \, .
  \end{equation*}
  
  In order to calculate the Lagrangian, we follow the Feng-Kurtz method outlined in \Sec\ref{Feng-Kurtz}. As a first step, we introduce the nonlinear generator
  \begin{align*}
   (H_n f)(x) &= \dfrac{1}{n} e^{-n f(x)} (\mathcal{Q}^n e^{n f})(x) = \nonumber \\
   & = n k (1 - x) \left( e^{n f(x + 1/n) - n f(x)} - 1 \right) + \nonumber \\
   &- n k x \left( e^{n f(x - 1/n) - n f(x)} - 1 \right)
  \end{align*}
  and calculate the limit
  \begin{multline*}
   (H_n f)(x) \to (H f)(x) = \\
   = k (1 - x) \left( e^{\di f(x)} - 1 \right) + k x \left( e^{- \di f(x)} - 1 \right) \, .
  \end{multline*}
  Indeed, the limit nonlinear generator $H$ depends on the function $f$ only through its first derivative. The next moves are to substitute $\xi$ for $\di f(x)$, defining the Hamiltonian
  \begin{equation}\label{HamiltonianCME}
   \mathcal{H}(x, \xi) := k (1 - x) \left( e^\xi - 1 \right) + k x \left( e^{-\xi} - 1 \right) \, ,
  \end{equation}
  and to compute the Legendre transform
  \begin{multline}\label{Lagrangian}
   \mathcal{L}(x, v) = k + v \sinh^{-1}\!\left(\dfrac{v}{\sqrt{4 k^2 x (1 - x)}}\right) + \\
    - \sqrt{k^2 + v^2 - \left[ k \left( 1 - 2x \right) \right]^2} - v \sinh^{-1}\!\left( \dfrac{k \left( 1 - 2x \right)}{\sqrt{4 k^2 x (1 - x)}} \right) \, .
  \end{multline}
  This Lagrangian is convex in $v$, and minimal with value zero at $v = k (1 - 2 x)$, which is the vector field of the deterministic dynamics \eqref{RRE}.
  
  For this simple system, we can explicitly compute the stationary distribution and verify that the process is in detailed balance with respect to it for any finite $n$. The distribution is a binomial one with parameters $n$ and $1/2$:
  \begin{equation}\label{invariantDistribution}
   \pi^n_x = \begin{pmatrix} n \\ nx \end{pmatrix} 2^{-n} \, .
  \end{equation}
  To obtain its large-deviation behavior for $n \to \infty$, the G\"artner-Ellis theorem \cite{hT09} gives
  \begin{equation*}
   \pi^n \asymp e^{n s} \qquad \text{as } n \to \infty \, ,
  \end{equation*}
  with
  \begin{equation}\label{S-GGF}
   s_x = - \left[ x \ln x + \left( 1 - x \right) \ln(1 - x) + \ln 2 \right] \, .
  \end{equation}
  
 \subsection{FDT and GGS for the reaction rate equation}\label{C}
  It is known that \Eq\eqref{RRE} has at least two generalized gradient structures \cite{aM11,mG10}, given the entropy \eqref{S-GGF}, which has derivative
  \begin{equation*}
   \di s_x = \ln\dfrac{1 - x}{x} \, .
  \end{equation*}
  The FDT singles out precisely one GGS, given the Hamiltonian \eqref{HamiltonianCME}. Indeed, by the properties of the Legendre transform, one can show that	
  \begin{align}
   \Psi^*_x(\xi) &= 2 \left[ \mathcal{H}\!\left(x, \dfrac{1}{2}(\xi - \di s_x) \right) - \mathcal{H}\!\left( x, -\dfrac{1}{2} \di s_x \right) \right] \label{Psi-H} \\
   &= 4 k \sqrt{x (1 - x)} \left( \cosh\dfrac{\xi}{2} - 1 \right) \, . \label{Psi-GGF}
  \end{align}
  The evolution equation,
  \begin{equation*}
   \dot{x}_t = \partial_\xi \Psi^*_{x_t}\!\big( \di s_{x_t} \big) \, ,
  \end{equation*}
  indeed reduces to the RRE \eqref{RRE}.
  
  Another possible choice for a GGS is given by a quadratic dissipation potential or, equivalently, by the friction matrix
  \begin{equation}\label{M}
   M_x = \dfrac{k \left( 1 - 2 x \right)}{\ln(1 - x) - \ln x} \, ,
  \end{equation}
  as shown, e.g., in \cite{aM11,OG97} and found in \cite{hcO15} by thermodynamic and geometric arguments.
  
  The GGS \eqref{Psi-GGF}-\eqref{S-GGF} was proposed in \cite{mG93} independently from any consideration of the large deviations of some underlying stochastic process, which is extremely remarkable. Arguments in favor of the quadratic dissipation potential associated to the friction matrix \eqref{M} are advanced in \cite{hcO15}.
  
 \subsection{Green-Kubo formula and diffusion approximations}\label{Green-Kubo}
  The scheme of \figurename\ref{fig:GFDT} provides the CME at level~$1^+$ with (i) the right deterministic limit (thus the right macroscopic dynamics), (ii) the right stationary distribution, and (iii) it gives an expression for the dissipation potential that can be computed by numerical simulations on level~$1$. Can these three features be reproduced with the method of \figurename\ref{fig:FDT} or, at least, by approximating the microscopic dynamics with a diffusion process? In this section we formulate an answer.
  
  If we use the scheme of \figurename\ref{fig:FDT}, at level $1^+$, we have the diffusion process that solves
  \begin{equation*}
   \di X^n_t = M^\text{GK}_{X^n_t} \di s_{X^n_t} \, \di t + \sqrt{\dfrac{2 M^\text{GK}_{X^n_t}}{n}} \diamond \di W_t \, ,
  \end{equation*}
  with the Green-Kubo formula
  \begin{equation}
   2 M^\text{GK}_x = n \lim\limits_{\tau \to 0} \dfrac{1}{\tau} \mathbb{E}\!\left[ \left( X^n_\tau - x \right)^2 \Big| X^n_0 = x \right] \, .
  \end{equation}
  Since we know that the CME is the correct model that approximates the $n$-particle process, we can calculate the theoretical result that we expect from simulations on level $1$:
  \begin{align}
   2 M^\text{GK}_x &= n \left[ (\mathcal{Q}^n f)(x) - 2 x (\mathcal{Q}^n g)(x) \right] = \nonumber \\
   &= k \left( 1 - x \right) + k x = k \label{M-CLE} \, ,
  \end{align}
  where $f(x) = x^2$, $g(x) = x$, and $\mathcal{Q}^n$ is the generator \eqref{generator} of the CME. We remark that $M^\text{GK}$ is, in general, a function of $x$, and it would be so if we had two distinct rate constants for the backward and forward reactions, $k^- \neq k^+$. With the entropy \eqref{S-GGF}, we get
  \begin{equation*}
   \di X^n_t = \dfrac{k}{2} \ln\dfrac{1 - X^n_t}{X^n_t} \, \di t + \sqrt{\dfrac{k}{n}} \, \di W_t \, ,
  \end{equation*}
  which has (ii) the right stationary distribution (by construction), (iii) a Green-Kubo expression for the friction matrix that can be computed by simulations on the microscopic level, but (i) the wrong drift and, therefore, the wrong deterministic limit (cf. the RRE \eqref{RRE}).
  
  Other possibilities of constructing a diffusion process for chemical reactions have been proposed: they aim to approximate the CME for a large number of particles. One of them is called \emph{chemical Langevin equation} \cite{dtG00}, which can also be thought of as the diffusion approximation \footnote{The diffusion approximation may be found by three heuristic arguments: the first corresponds to expanding the generator to first order in~$1/n$, the second to considering the Kramers-Moyal expansion in the equation for the law, and the third to replacing the Poisson noise, which describes the jumps, by a Brownian one for large~$n$ \cite{ngvK83,EK05,KKP14,jmH15}.} of the CME, and reads
  \begin{align}
   \di Y^n_t &= k (1 - 2 Y^n_t) \, \di t + \sqrt{\dfrac{k}{n}} \, \di W_t \label{CLE} \\
   &= M^\text{GK}_{Y^n_t} \di s^\text{CLE}_{Y^n_t} \, \di t + \sqrt{\dfrac{2 M^\text{GK}_{Y^n_t}}{n}} \diamond \di W_t \nonumber \, ,
  \end{align}
  with the entropy
  \begin{equation}
   s^\text{CLE}_y = - 2 \left( y - \dfrac{1}{2} \right)^2
  \end{equation}
  and the friction matrix \eqref{M-CLE}. This process has (i) the right deterministic limit, (iii) a Green-Kubo expression for the friction matrix, but (ii) the wrong entropy, viz., the wrong stationary distribution.
  
  A third choice is the \emph{log-mean equation} \cite{BBGBD15}
  \begin{align*}
   \di Z^n_t &= M_{Z^n_t} \di s_{Z^n_t} \, \di t + \sqrt{\dfrac{2 M_{Z^n_t}}{n}} \diamond \di W_t \nonumber \\
   &= k (1 - 2 Z^n_t) \, \di t + \sqrt{\dfrac{2}{n} \dfrac{k \left( 1 - 2 Z^n_t \right)}{\ln(1 - Z^n_t) - \ln Z^n_t}} \diamond \di W_t \, ,
  \end{align*}
  with the entropy \eqref{S-GGF} and the friction matrix \eqref{M}. It has (i) the right deterministic limit, (ii) the right stationary distribution, but (iii) the friction matrix cannot be derived by the Green-Kubo formula.
  
  Among the three diffusion processes, only one is consistent with the CME for relatively small deviations from the deterministic limit. Indeed, the large-deviation Lagrangian of the CLE \eqref{CLE},
  \begin{equation}\label{DiffusionApproximation}
   \mathcal{L}_\mathrm{D}(x, v) = \dfrac{1}{2 k} \left[ v - k(1 - 2x) \right]^2 \, ,
  \end{equation}
  is the quadratic form in $v$ that approximates the Lagrangian \eqref{Lagrangian} around the deterministic vector field \eqref{RRE} (cf.~\figurename\ref{fig:Hamiltonian}). This further justifies why the CLE is \emph{the} diffusion approximation of the CME.
  
  In conclusion, there is no way for a diffusion process to satisfy the requirements (i)-(iii) simultaneously. In order to satisfy all three requirements, one should move to more general Markov processes and to non-quadratic dissipation potentials, as indicated in \figurename\ref{fig:GFDT}.

 \subsection{Numerical experiments}\label{numerics}
  The goal of the scheme depicted in \figurename\ref{fig:GFDT} is to infer the structure of a more macroscopic level of description from a more microscopic one, which in this case is represented by the Brownian motion of many independent particles in a double-well potential. The latter, for $\epsilon$ small enough, is well approximated by the CME. The approach advanced in this paper, thus, suggests the following procedure: we perform simulations on the microscopic level, use the coarse-graining map \eqref{empiricalProcess}, and approximately compute the large-deviation rate functions for the CME, which give us the GGS of the deterministic limit.
  
  Since, for this simple problem, we know everything analytically, we can compare the numerical results with the exact ones. In particular, we compare the values of the reaction constant obtained by simulation with the one provided by Kramers' formula \cite{haK40,BdH15}
  \begin{equation}\label{Kramers}
   \bar{k} = \dfrac{\sqrt{- V''(0) \, V''(A)}}{2 \pi} e^{\left( V(A) - V(0) \right)/\epsilon} \, .
  \end{equation}
  
  In accordance with the generalized FDT, we should estimate the entropy function by looking at the large deviations of the stationary distribution of the process. For simplicity, however, here we suppose to know the entropy and concentrate on the dynamic large deviations. The latter task is, in general, a very hard one, since the dynamic rate function takes values in a space of curves; in other words, we are dealing with a very high-dimensional problem. However, the form \eqref{rateFunction} of the rate function for time-homogeneous Markov processes in terms of a Lagrangian, a local function of just two variables, notably reduces the issue of dimensionality. For this reason, we think that the Feng-Kurtz method, whose main scope is to find the Lagrangian by studying the convergence of nonlinear generators, represents an excellent framework for our purposes. Hence, we shall develop a numerical implementation of the Feng-Kurtz method introduced in \Sec\ref{Feng-Kurtz}.
  
  The core of the Feng-Kurtz method is to calculate the nonlinear generator
  \begin{equation}\label{nonlinearGenerator}
   (H^n f)(x) := \dfrac{1}{n} e^{-n f(x)} \lim\limits_{\tau \to 0} \dfrac{\mathbb{E}\!\left[ e^{n f(X^n_\tau)} \big| X^n_0 = x \right] - e^{n f(x)}}{\tau} \, ,
  \end{equation}
  whose limit defines the Hamiltonian
  \begin{equation}\label{HamiltonianNum}
   (H^n f)(x) \to \mathcal{H}(x, \di f(x)) \quad \text{as } n \to \infty \, ,
  \end{equation}
  which here has to be understood in a purely numerical sense: the number of particles $n$ should be large enough for an acceptable accuracy on $\mathcal{H}$. It is not the purpose of this paper to go into the details of the control of this convergence, but only to show the successfulness of this method in the context of the present example. In the same way, for the moment we do not have the goal of constructing efficient simulations, nor to pursue any statistical rigor, which we reserve for future work. For instance, in the same spirit of inference for generators of continuous-time Markov processes \cite{BS05}, a theory of inference for nonlinear generators should be developed.
  
  Since the limit nonlinear generator \eqref{HamiltonianNum}, if it exists, depends only on $x$ and $\di f(x)$, it is sufficient to consider linear functions.
  
  The numerical discretization of \Eq\eqref{nonlinearGenerator} contains five parameters that, in principle, we would like to take to their limits, but in our numerical experiment attain only finite values:
  \begin{itemize}
   \item The noise intensity $\epsilon \to 0$, which controls the separation of time scales. Kramers mentions in his seminal paper \cite{haK40} that $\epsilon = 0.2$ is sufficient: the process becomes approximately Markov and is well described by the CME. We actually work with $\epsilon = 0.15$. In applications, however, this is not a parameter, but a model datum.
   \item The time-step size $\Delta t \to 0$ of the numerical scheme used to simulate the SDE \eqref{Kramers-SDE}, which should resolve the microscopic dynamics, guarantee stability of the scheme, and be smaller than the local equilibration time $t_2$ \footnote{An estimation of the equilibration time $t_2$ should be done separately, for instance with the help of the various methods available in the literature \cite{DLLN16,afV98,BLS15}. In our case, the ratio $t_2/\Delta t$ is of the order of $10$.}. We consider the numerical value $\Delta t = 0.01$.
   \item The time interval $\tau \to 0$. This time constant should be ``macroscopically small'', i.e., much smaller than the typical jump time $t_1 = k^{-1}$, but also larger than the equilibration time $t_2$, in such a way that we retain only the macroscopic features of the process~$X^n$, namely, we neglect the underlying diffusive nature of it. We take $\tau = \bar{k}^{-1}/50$, where $\bar{k}$ is given by formula \eqref{Kramers}. In more general contexts, we may not know the values of the characteristic times in advance and should perform an appropriate estimation of them.
   \item The number of particles $n \to \infty$. Since the particles are independent and the observables $f$ are linear, the nonlinear generator loses its dependence on $n$. This parameter, then, only controls the discretization of the space $\mathcal{X}^n$, and $n = 6$ is enough for our purposes.
   \item The sample size $N \to \infty$ over which we calculate the expectation. To obtain good statistics, enough jumps in the time interval $\tau$ should occur. Since the average jump time is $t_1$, we need $N \tau \gg t_1$, and we choose $N = 10^5$.
  \end{itemize}
  We have built the chain of inequalities
  \begin{equation*}
   \Delta t \ll t_2 \ll \tau \ll t_1 \ll N \tau \, .
  \end{equation*}
  For definiteness, we choose the quartic potential
  \begin{equation*}
   V(q) = \left( q^2 - 1 \right)^2 \, .
  \end{equation*}
  
  \begin{figure*}[t]
   \begin{minipage}[t]{.482\textwidth}
    \centering
    \includegraphics[width=1.05\linewidth]{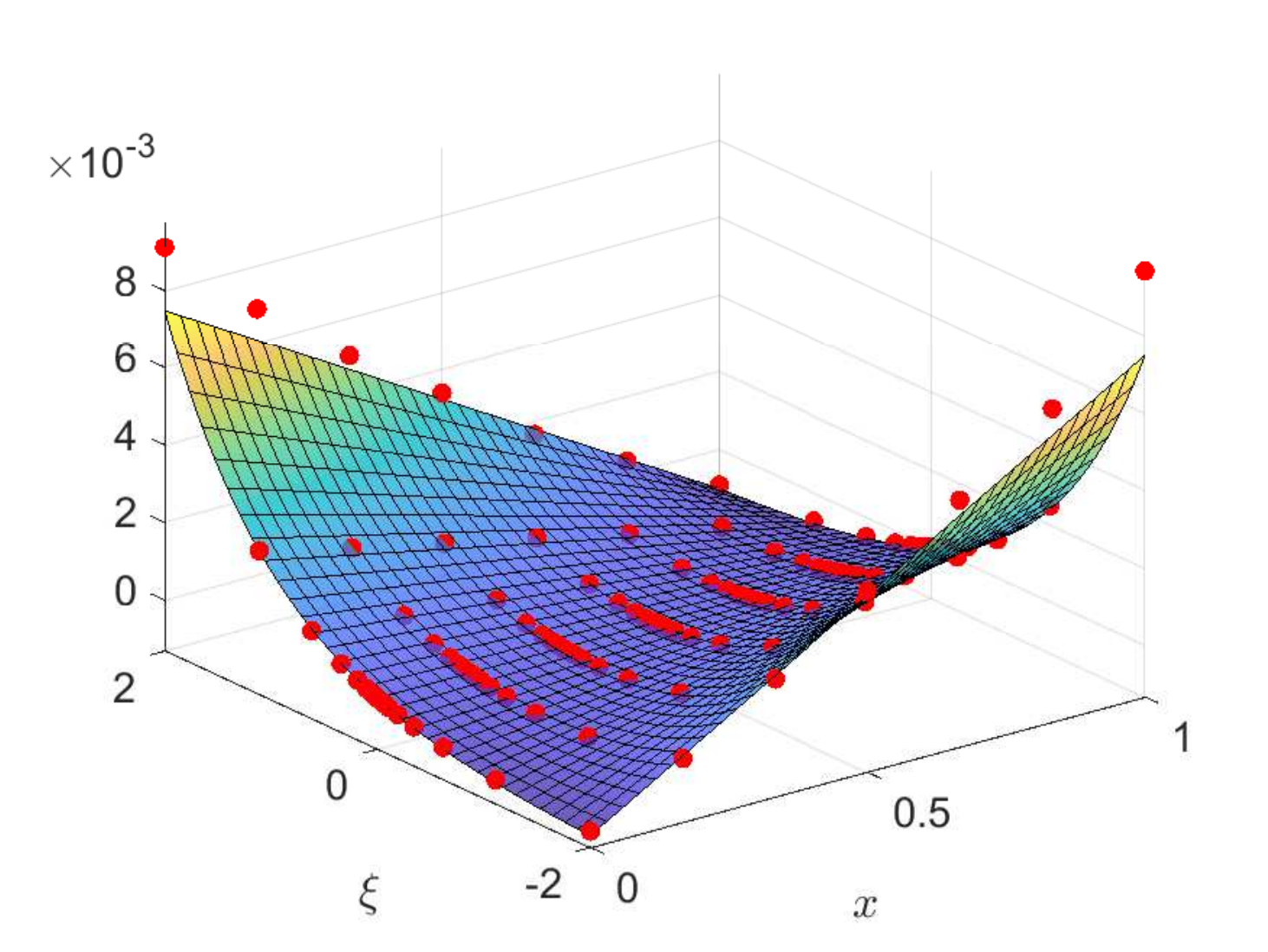}
  	\caption{Numerical estimate of the Hamiltonian compared with the theoretical result. The latter is represented by the smooth surface, and the former by the red dots. The values of the parameters are: $\epsilon = 0.15$, $\Delta t = 10^{-2}$, $\tau = \bar{k}^{-1}/50$, $n = 6$, $N = 10^5$, $\xi_\text{max} = 2$.}
  	\label{fig:Hamiltonian}
   \end{minipage}\quad \ \ \
   \begin{minipage}[t]{.482\textwidth}
  	\centering
  	\includegraphics[width=.82\linewidth]{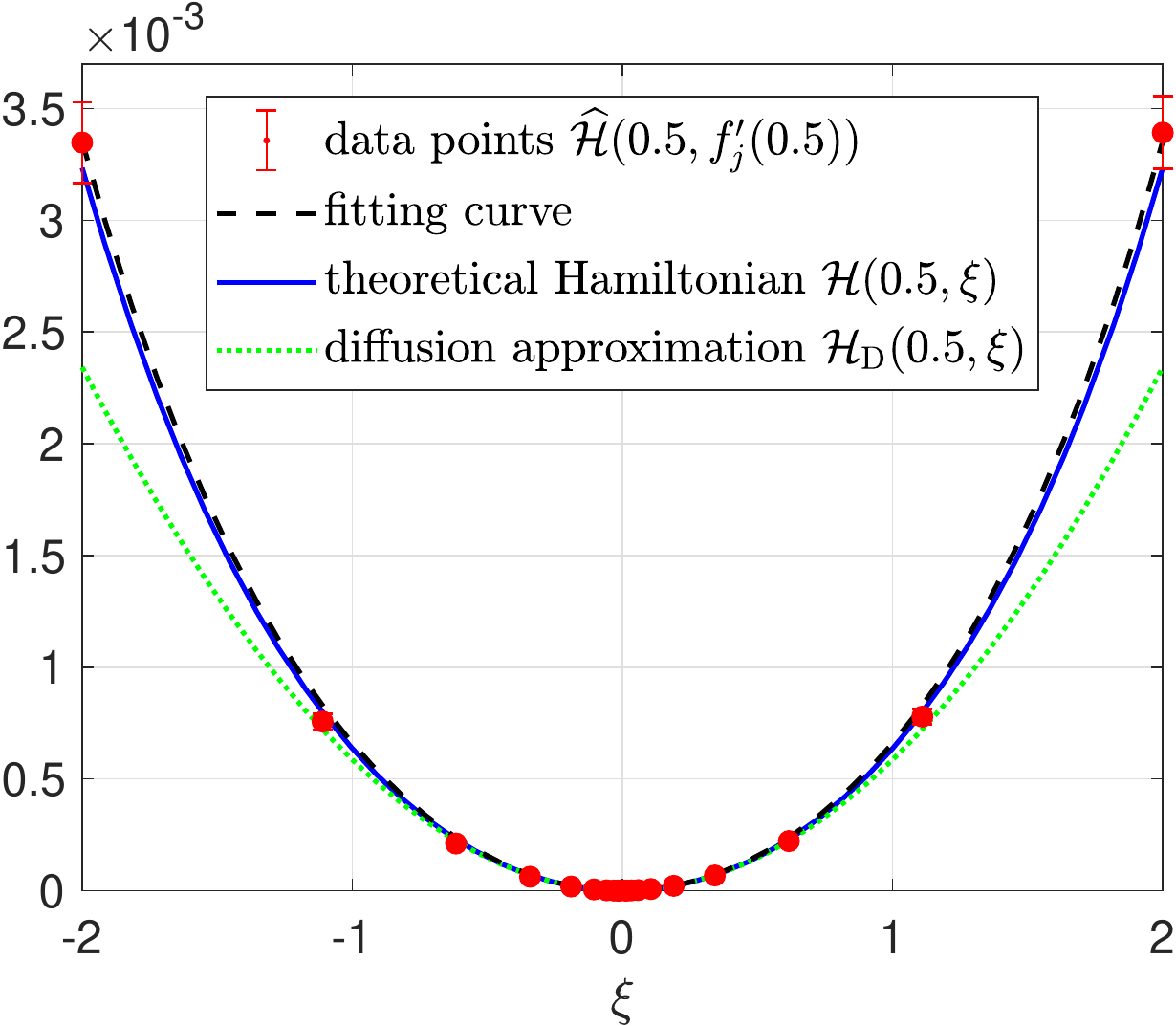}
  	\caption{Fitting of the data points $\widehat{\mathcal{H}}(0.5, f_j'(0.5))$ with the function $k/2 \left( e^\xi - 1 \right) + k/2 \left( e^{-\xi} - 1 \right)$ (\Eq\eqref{HamiltonianCME}) and comparison to the theoretical Hamiltonian and its diffusion approximation \eqref{DiffusionApproximationHamiltonian}, both with $k = \bar{k}$. The confidence intervals are at level 99\%.}
  	\label{fig:fitting}
   \end{minipage}
  \end{figure*}
  
  The numerical setup is the following.
  \begin{enumerate}
   \item We select the observables $f_j(x) = \xi_j x$, with the $\xi_j$ logarithmically spaced \footnote{We choose a small value $\xi^+_1$. The positive values $\xi^+_j$ are in geometric progression starting from $\xi^+_1$, and the negative ones are $\xi^-_j = - \xi^+_j$.} in the interval $[-\xi_\text{max}, \xi_\text{max}]$ with $\xi_\text{max} = 2$. The logarithmic spacing has the aim of resolving the region around $\xi = 0$ sufficiently well.
   \item For every $x \in \mathcal{X}^n$, we run $N$ simulations, of length $\tau$, of the $n$ independent SDEs \eqref{Kramers-SDE}. We use an Euler-Mayurama scheme with step size $\Delta t$. The starting point for $n x$ particles is $A$, and for the others is $B$. We need not care about equilibration in the wells because $\tau \gg t_2$.
   \item We index the simulations by $k$ and say that the random variable $X_\tau$ has $x^k_\tau$ as its realization. After the $k$-th simulation started from $x$, we compute the quantities
   \begin{equation}\label{sample}
    h_j^k(x) := e^{n f_j(x^k_\tau)} \, .
   \end{equation}
   There is one such quantity for each $x$, $j$ and $k$.
   \item Since the sample is automatically extracted from the (approximate, because of time discretization) law of the process, to estimate the expectation in \Eq\eqref{nonlinearGenerator} we only need to take the simple averages
   \begin{equation}
    \dfrac{1}{N} \sum\limits_{k = 1}^N h_j^k(x) \, .
   \end{equation}
   Therefore, our estimator for the nonlinear generator is
   \begin{equation}
    (\widehat{H} f_j)(x) = \dfrac{1}{n \tau} \left( e^{-n f_j(x)} \dfrac{1}{N} \sum\limits_{k = 1}^N h_j^k(x) - 1 \right) \, .
   \end{equation}
  \end{enumerate}
  Then, from the Feng-Kurtz theory, we know that computing the nonlinear generator on an observable $f_j$ corresponds to the estimate of the Hamiltonian
  \begin{equation}
   \widehat{\mathcal{H}}(x, \di f_j(x)) = (\widehat{H} f_j)(x) \, .
  \end{equation}
  
  This algorithm provides us with an estimate at the points~$(x, \xi_j) \in \mathbb{R}^2$. The results are displayed in \figurename\ref{fig:Hamiltonian}, where the solid surface is the Hamiltonian \eqref{HamiltonianCME} with ${k = \bar{k}}$, and the red dots are its estimated values, which show excellent agreement.
  
  If we assume the functional form of the Hamiltonian, we are able to determine the whole function in $\mathbb{R}^2$ by using a finite set of observables. For Markov jump processes on a graph, we can always expect the Hamiltonian to be of the form \cite{MPPR17}
  \begin{equation}
   \mathcal{H}(x, \xi) = \sum\limits_{\nu} r_\nu(x) \left( e^{\xi \cdot \nu} - 1 \right) \, ,
  \end{equation}
  where $\nu$ represent the directed path between two nodes (states), and the sum is over all paths. In our one-dimensional case, this is reduced to the determination of the functions $r_1$ and $r_2$ in
  \begin{equation}\label{Hamiltonian}
   \mathcal{H}(x, \xi) = r_1(x) \left( e^\xi - 1 \right) + r_2(x) \left( e^{-\xi} - 1 \right) \, .
  \end{equation}
  The coefficients $1$ and $-1$ in the exponents represent the stoichiometric coefficients of the two reaction paths (forward and backward) and are generalizable to any reaction network \cite{MPPR17}.
  
  If, in addition, we know the functional expressions for the rates, we may determine the reaction constants by fitting the simulation points with the function \eqref{HamiltonianCME}. We do so with the values $(\widehat{H} f_j)(x)$ at $x = 0.5$, a cross section of \figurename\ref{fig:Hamiltonian}. In \figurename\ref{fig:fitting}, we compare this fitting procedure to the theoretical Hamiltonian with $k = \bar{k}$. In addition, we provide the comparison with the prediction of the diffusion approximation (cf.~\Sec\ref{Green-Kubo}), namely, with the Legendre transform of the Lagrangian \eqref{DiffusionApproximation},
  \begin{equation}\label{DiffusionApproximationHamiltonian}
   \mathcal{H}_\mathrm{D}(x, \xi) = \dfrac{k}{2} \xi^2 + k \left( 1 - 2 x \right) \xi 	\, .
  \end{equation}
    
\begin{figure}[h]
   \centering
   \includegraphics[width=1.05\linewidth]{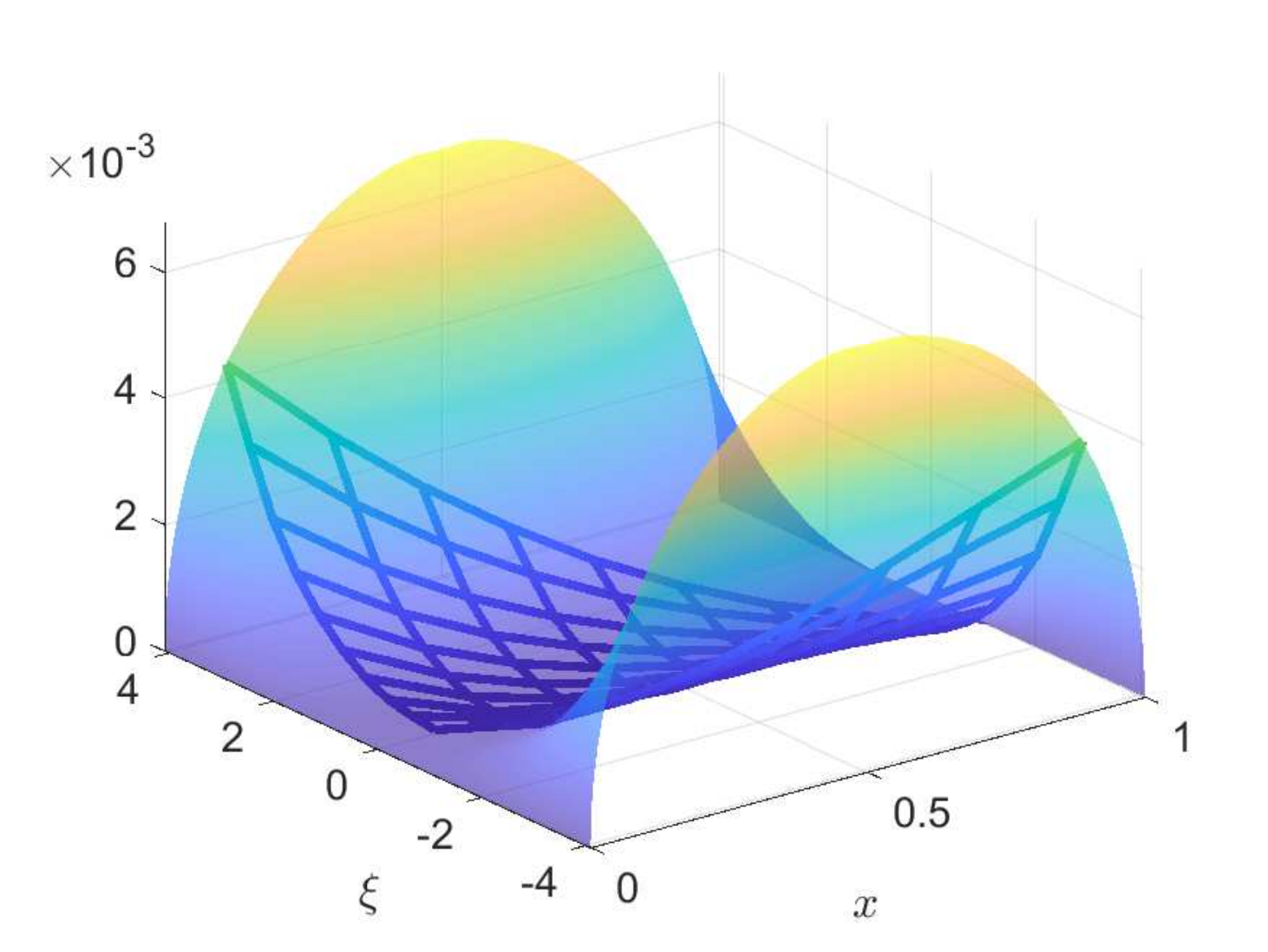}
   \caption{Numerical estimate of the dissipation potential in the region $\left\{ 0 \leq x \leq 1, -4 + \di s_x \leq \xi \leq 4 + \di s_x \right\}$ compared with the theoretical result. The latter is represented by the smooth surface, and the former by the mesh, which was found by interpolation of the simulated Hamiltonian and evaluation with the MATLAB function \texttt{scatteredInterpolant}. The error is everywhere smaller than $1.4 \cdot 10^{-4}$.}
   \label{fig:DissipationPotential}
  \end{figure}

  In order to find the dissipation potential, we assume to know the entropy function \eqref{S-GGF}. According to \Eq\eqref{Psi-H}, we would like to compute
  \begin{equation}
   \widehat{\Psi}^*_x(\xi) = 2 \left[ \widehat{\mathcal{H}}\!\left(x, \dfrac{1}{2}(\xi - \di s_x) \right) - \widehat{\mathcal{H}}\!\left( x, -\dfrac{1}{2} \di s_x \right) \right] \, .
  \end{equation}
  Since we know the values of the Hamiltonian only at discrete points, first we need to interpolate it. Via the MATLAB function \texttt{scatteredInterpolant} \footnote{The function \texttt{scatteredInterpolant} uses a Delaunay triangulation of the scattered sample points to perform interpolation \cite{iA02}. The resulting function can then be evaluated at query points.}, we obtain an interpolation for $\widehat{\Psi}^*$ in the region $\left\{ 0 \leq x \leq 1, - 2 \xi_\text{max} + \di s_x \leq \xi \leq 2 \xi_\text{max} + \di s_x \right\}$. Disregarding any statistical consideration again, the result is shown in \figurename\ref{fig:DissipationPotential}.
  
  In this very simple example of a chemical reaction, coarse-graining means estimating the reaction constant, namely the parameter $k$ in the linear generator \eqref{generator}. Consequently, the limit nonlinear generator or the Hamiltonian \eqref{Hamiltonian} does not really contain different information than the linear generator. Hence, in the present context, one could just estimate the transition rates by one of the numerous methods already available in the literature (e.g., \cite{eVE06,FVEW15,HBSBS14}). Moreover, considering more general reaction networks, typical problems of standard approaches remain: high local minima of the potential landscape are rarely explored, boundaries between macroscopic states are not easy to set, the recrossing problem is typical \cite{FVEW15}. The approach proposed here presents the additional complication of being ``stiff'' because of the strong nonlinearities in \Eq\eqref{sample}.
  
  However, we emphasize that our aim is to directly resolve the full structure \eqref{Psi-GGF}, and not just the rates: this is not a feature of already available methods. We also expect that our viewpoint will constitute an advantageous tool for less trivial systems, where the convergence of the nonlinear generator carries substantial new information, such as problems of homogenization \cite{MS13} or systems where the structure of the macroscopic dynamics is not known in advance \cite{BC12}.
  
  The last main issue is numerical efficiency. Many known numerical methods in statistical mechanics \cite{hT11} are based on \emph{importance sampling}: the probability distribution of a random variable is changed, often by ``exponential tilting'', in such a way that rare events become less rare and can more easily be observed. Physically, this corresponds to biasing the system by an external force. Following the ideas in \cite{HBSBS14}, we would like to develop biased methods for nonlinear generators.

\section{Conclusions\label{sec:conclusions}}
 We have considered the statistical mechanics of a physical system with many degrees of freedom represented by a parameter $n$. In particular, we have studied the structural properties of the deterministic limit ($n \to \infty$) and the fluctuation properties around that limit. The right mathematical framework for this task is the theory of large deviations.
  
 Classically, a fluctuation-dissipation theorem of the second kind is formulated for a diffusion process: in the limit of large but finite $n$, it gives the definition of the friction properties of a system, which characterize its most probable evolution, in terms of its fluctuation properties around this evolution. In this paper, we have extended this definition to more general Markov processes with detailed balance.
 
 The friction properties are encoded in a nonlinear generalization of a gradient flow, called \emph{generalized gradient flow}, which has the following feature: the driving function, interpreted as the relevant thermodynamic potential, characterizes equilibrium and never decreases along the evolution. For example, for closed systems, the driving function is the thermodynamic entropy. The generalized gradient flow is entirely determined by the driving function and a dissipation potential.
 
 The fluctuation properties are characterized by two large-deviation principles: a static one, whose rate function contains the information on the equilibrium states, hence it is the relevant thermodynamic potential; and a dynamic (pathwise) one, where the rate function describes the deviations of the stochastic trajectories from the most probable one, namely the macroscopic evolution, in the limit $n \to \infty$.
 
 The generalized FDT establishes the definition of friction in terms of the fluctuations: the dissipation potential, which characterizes friction, is uniquely defined by the dynamic rate function, which describes fluctuations.
 
 The most important consequence of the generalization is an extended theory of coarse-graining. If the classical theory of fluctuations allowed us to handle only diffusion processes -- the typical setting of macroscopic ``hydrodynamic-like'' equations and Green-Kubo relations --, now the class of systems includes ``rare-event-like'' systems, which are much better described by jump processes rather than diffusions.
 
 In this context, we have tested the new method in the example of a simple model of a monomolecular chemical reaction, the Kramers escape problem. Although this elementary illustration has been proven successful, it certainly requires refinement from both the standpoint of the statistical solidity and of the efficiency of the algorithm, especially because we aim to apply the method to more complex systems, such as plasticity and the dynamics of glasses.
 
 The connection between large deviations and generalized gradient flows is restricted to purely dissipative systems. Indeed, its extension to dynamics with a non-dissipative component has not been established yet, although a few attempts \cite{DPZ13,KLMP18} have been made. We expect that the general class of FDTs should hold for evolutions equations of the GENERIC type, where the non-dissipative component is modeled by a Poisson structure and the dissipative one by a generalized gradient structure.
 
\begin{acknowledgments}
 We are grateful to Mohsen Talebi, Aleksandar Donev, Patrick Ilg, Robert Riggleman, and Elijah Thimsen, who considerably helped us to improve our understanding and the presentation of the ideas.
\end{acknowledgments}

\bibliographystyle{abbrvplain}

\end{document}